\documentclass[manuscript]{aastex}

\usepackage{natbib}  % added by simon 3/7/08
\usepackage{epsfig,graphicx,graphics}

\usepackage[latin1]{inputenc}

\shorttitle{Cavity clearing in HD~142527}
\shortauthors{Casassus et al.}

\slugcomment{ Based on observations obtained at the Gemini
  Observatory, which is operated by the Association of Universities
  for Research in Astronomy, Inc., under a cooperative agreement with
  the NSF on behalf of the Gemini partnership: the National Science
  Foundation (United States), the Science and Technology Facilities
  Council (United Kingdom), the National Research Council (Canada),
  CONICYT (Chile), the Australian Research Council (Australia),
  Minist\'{e}rio da Ci\^{e}ncia e Tecnologia (Brazil) and Ministerio
  de Ciencia, Tecnolog\'{i}a e Innovaci\'{o}n Productiva
  (Argentina). The Gemini run ID is GS-2011A-Q-27.  Also based on
  based on observations made with ESO Telescopes at the Paranal
  Observatory under DDT program 287.C-5041(A).}

\begin{document}

\title{The dynamically disrupted gap in  HD~142527}

\author{S. Casassus\altaffilmark{1}, S. Perez M.\altaffilmark{1},
  A. Jord\'an\altaffilmark{2}, F. M\'enard\altaffilmark{3,4}, J.
  Cuadra\altaffilmark{2}, M. R. Schreiber\altaffilmark{5}, 
%Alberto  Rebassa-Mansergas\altaffilmark{5}, Gisela Romero\altaffilmark{5},
  A.~S. Hales\altaffilmark{6}, B. Ercolano\altaffilmark{7}}

\altaffiltext{1}{Departamento de Astronom\'{\i}a, Universidad de Chile}
\altaffiltext{2}{Departamento de Astronom\'ia y Astrof\'isica,
Pontificia Universidad Cat\'olica de Chile, 7820436 Macul, Santiago,
Chile}
\altaffiltext{3}{UMI-FCA 3386, CNRS / INSU, France and Departamento de Astronom\'{\i}a, Universidad de Chile}
\altaffiltext{4}{Institut de Plan\'etologie et d'Astrophysique de Grenoble (IPAG) CNRS/UJF UMR 5274, Grenoble, France}
\altaffiltext{5}{Departamento de F\'{\i}sica y Astronom\'{\i}a, Universidad Valparaiso, Av. Gran
Bretana 111, Valparaiso, Chile.}
\altaffiltext{6}{Joint ALMA Observatory, Alonso de C\'ordova 3107, Vitacura 763-0355, Santiago - Chile}
\altaffiltext{7}{University Observatory, Ludwig-Maximillians University, Munich}

\begin{abstract}

The vestiges of planet formation have been observed in debris disks
harboring young and massive gaseous giants. The process of giant
planet formation is terminated by the dissipation of gas in the
protoplanetary disk. The gas-rich disk around HD~142527 features a
small inner disk, a large gap from $\sim$10 to $\sim$140~AU, and a
massive outer disk extending out to $\sim$300~AU. The gap could have
been carved-out by a giant planet.  We have imaged the outer regions
of this gap using the adaptive-optics camera NICI on Gemini South. Our
images reveal that the disk is dynamically perturbed. The outer
boundary of the roughly elliptical gap appears to be composed of
several segments of spiral arms.  The stellar position is offset by
$0.17\pm0.02^{\prime\prime}$ from the centroid of the cavity,
consistent with earlier imaging at coarser resolutions.  These
transient morphological features are expected in the context of disk
evolution in the presence of a perturbing body located inside the
cavity.  We perform hydro-dynamical simulations of the dynamical
clearing of a gap in a disk.  A 10~M$_\mathrm{jup}$ body in a circular
orbit at $r = 90$~AU, perturbs the whole disks, even after thousands
of orbits. By then the model disk has an eccentric and irregular
cavity, flanked by tightly wound spiral arms, but it is still evolving
far from steady state.  A particular transient configuration that is a
qualitative match to HD~142527 is seen at 1.7~Myr.

\end{abstract}


\section{Introduction}

The lifetime of disks sets the time available for the planet formation
process.  In the Solar System, chronology of the oldest solid
components based on radionuclides (e.g., Al$^{26}$) indicates that the
duration timescale of the Solar Nebula was short, of order 2--3~Myr
\citep{2006EM&P...98...39M}. Direct imaging of a
$9\pm3$~M$_\mathrm{jup}$ planet \citep{2010Sci...329...57L} inside the
debris disk of $\beta$~Pic proves that large gas giants can form by
12$^{+8}_{-4}$~Myr.  In protostellar systems, by 5-10~Myr no gas is
left \citep{2006ApJ...651.1177PF}. The evidence points at early giant
planet formation, as in the candidate massive protoplanets LkCa~15~b
\citep{2012ApJ...745....5K} and T~Cha~b
\citep{2011A&A...528L...7H} (although both remain to be
  confirmed, e.g. by direct imaging).

Radial gaps in gas-rich protoplanetary disks are currently thought of
as possible signposts of planet formation. Previous $H$ and $K$ band
coronographic images of HD~142527 reveal a hole in the disk, about
100~AU in radius \citep{2006ApJ...636L.153F}.  This inner cavity is in
fact a gap with an inner boundary, abutting on an inner disk
\citep{2004Natur.432..479V}, roughly 10~AU in radius (Anthonioz et
al., in preparation), and up to 30~AU~\citep{2011A&A...528A..91V},
that accounts for the large near-IR excess of HD~142527
\citep{2010PASJ...62..347F}. Radiative transfer modelling of the
spectral-energy distribution (SED) and NIR images are consistent with
an inclination angle of $\sim$20~deg \citep[close to
  face-on,][]{2011A&A...528A..91V}.

The youth of HD~142527A, at $\sim 140$~pc\footnote{Assuming it is
  associated to Sco~OB2; the Hipparcos parallax of
  4.29$\pm$0.98~marcsec lower-limits its distance to 138~pc, at
  3~$\sigma$.}, is evident from the copious amounts of gas
\citep{2011ApJ...734...98O} in its $\sim$0.1~M$_\odot$ circumstellar
disk \citep{2011A&A...528A..91V}. \citet{2006ApJ...636L.153F} estimate
a stellar age of $2^{+2}_{-1}~$Myr and a mass of
1.9$\pm$0.3~M$_\odot$.

%We constrained basic
%stellar parameters from echelle spectroscopy ***** : its effective
%temperature is $T_\mathrm{eff} = 6400\pm200$, its surface gravity is
%$\log(g) = 3.5\pm0.05$, and its spectrum is extinct by $A_V =
%1.7\pm0.2$.  Accordingly, the bolometric luminosity of HD~142527A is
%$L=36\pm1~L_\odot$. Pre-main sequence evolutionary models
%\citep{2000A&A...358..593S} place HD~142527A at $2\pm1$~Myr, with a
%mass of $2.7\pm0.2~M_\odot$. These numbers improve on the precision of
%previous work \citep{2006ApJ...636L.153F}.  ****PERHAPS MOVE TO SEC 2
%*****
%

Here we report on deep adaptive optics near-IR imaging of the disk and gap of HD~142527.
% which reveal structure in the outer edge of the central cavity. 
%We interpret this structure in terms of dynamical clearing by a planetary-mass companion. 
\S~\ref{sec:observations} describes our observational setups,
\S~\ref{sec:obsresults} emphasizes our main observational results,
\S~\ref{sec:analysis} interprets, and \S~\ref{sec:conclusion} concludes.

\section{Observations} \label{sec:observations}

%\subsection{NICI adaptive optics imaging}

\begin{figure}
\begin{center}
\includegraphics[width=0.8\textwidth,height=!]{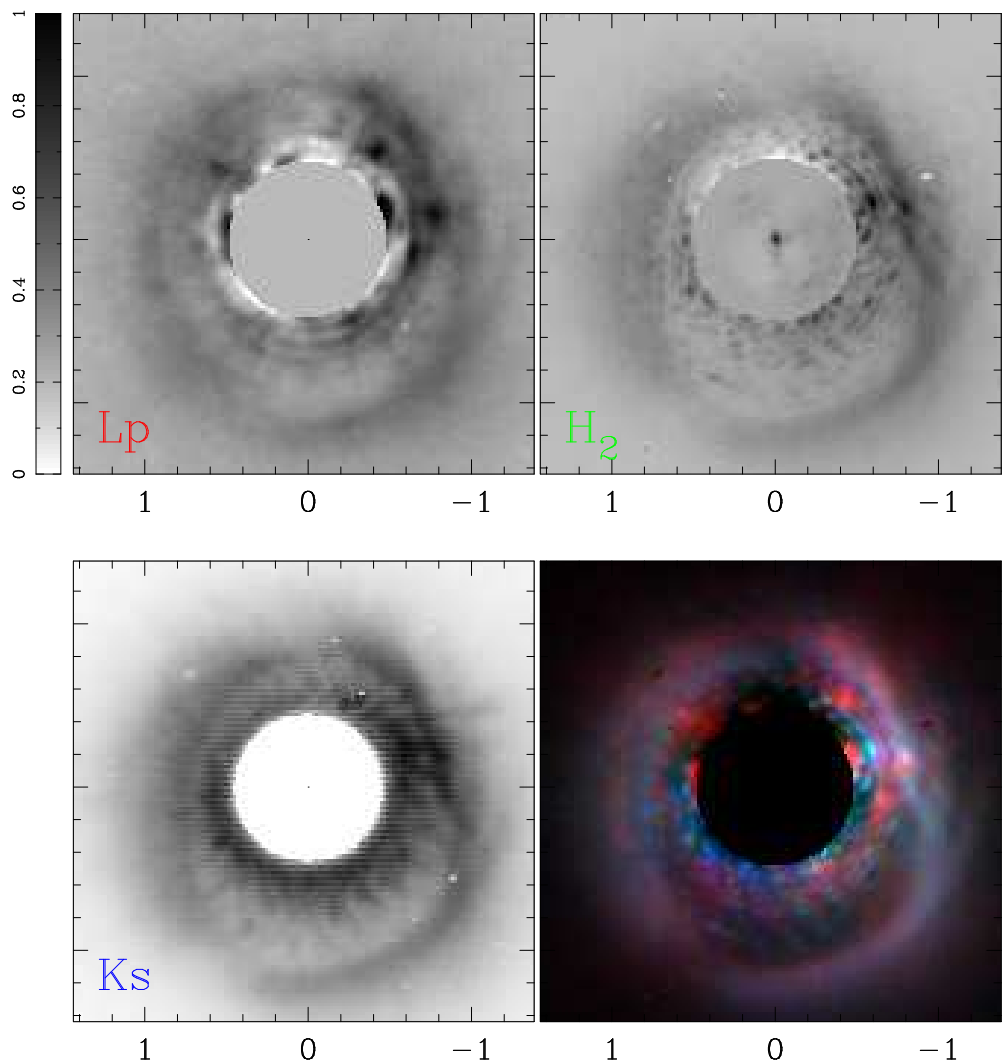}
\end{center}
\caption{ Infrared adaptive-optics imaging of HD142527 and its
  protoplanetary disks, from 11-Jun-2011. In x and y we indicate
  offset J2000 RA \& Dec, in arcsec (so North is up and East is to the
  left).  From left to right we show reduced images obtained in $L_p$,
  H$_2$(1-0)S(1), and $K_s$, and an RGB combination following the
  color codes indicated in each image. The linear gray scales are
  normalised over the entire range of intensity values outside a
  radius of 0.45~arcsec, containing the core of the diffracted stellar
  light, and abutting on the halo of the stellar glare. Inside this
  region the images are masked, except for the H$_2$ image, where the
  stellar PSF has been attenuated with azimuthal averages.  A roughly
  elliptical outer ring is clearly seen. The star is offset from the
  center of the cavity by 0.17$\pm$0.02~arcsec.
 \label{fig:NICI}}
\end{figure}

{\em NICI adaptive optics imaging}. We conducted IR adaptive optics
observations of HD~142527 with the NICI coronograph
\citep{2010ApJ...720L..82B} on the Gemini telescope.  NICI is a
dual-channel imager, the images were acquired simultaneously, by
pairs, a different filter in each channel. Here we report on two pairs
of filters: 1- broadband $Ks$ (2.15~$\mu$m) with H$_2$(1-0)S(1)
(2.12~$\mu$m), both with a 0.22$^{\prime\prime}$ mask occulting 95\%
of the stellar light, and both exposed to 2~s and 30~s, for a total of
1~h; 2- broadband $L'$ (3.8~$\mu$m) with Br$\gamma$ (2.17~$\mu$m),
both maskless and exposed to 0.38s, for a total of 1~h.  All setups
were acquired in stare mode, compensating for parallactic angle
rotation. The result of our NICI imaging is shown in
Fig.~\ref{fig:NICI}, after PSF subtraction\footnote{we used a Moffat
  profile fit to the wings of the PSF}.

%\subsection{SINFONI integral field spectra}

{\em SINFONI integral field spectra}. We also conducted complementary
integral-field-unit spectroscopy with the SINFONI instrument on the
VLT in order to search for possible planetary objects in the gap. In
our setup we chose a very narrow jittering throw that samples the
inner $0.45\times0.45$~arcsec$^{-2}$ only. We processed the ESO
pipeline datacubes using spectral deconvolution techniques
\citep{2007MNRAS.378.1229T}, and also through conventional PSF
subtraction (using a PSF standard star). The detection limits we
achieve are consistent with those achieved by
\citet{2007MNRAS.378.1229T} in AB~Dor with the same instrument
($\sim$9~mag contrast at 0.2~$^{\prime\prime}$ separation).
%We note that further SINFONI observations are underway to 
%sample the whole cavity.

\section{Observational results} \label{sec:obsresults}

\subsection{Size and morphology of the cavity and outer disk inner rim}

%AJ: previous phrasing of first sentence was putting our interpretation forward
%    the image highlight is independent of our interpretation
% FMe, nice, needed to to that

\begin{figure}
\begin{center}
\includegraphics[width=\textwidth,height=!]{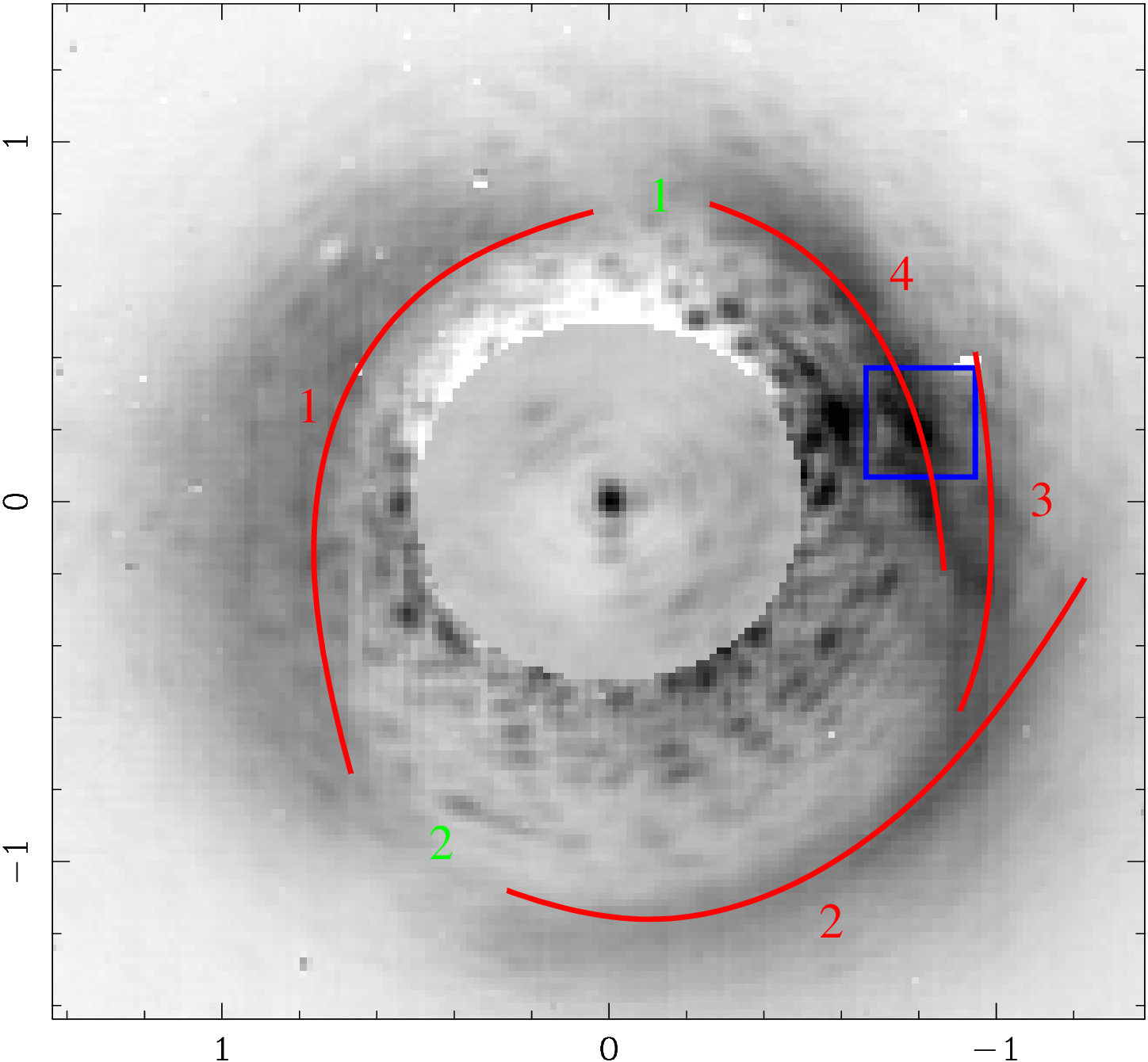}
\end{center}
\caption{ Annotations on the H$_2$(1-0)S(1) image also show on
  Fig.~\ref{fig:NICI}. In red we indicate the spiral arms we could
  identify, with red identification numbers. The green numbers
  indicate the positions of the two intensity nulls, or gaps, along
  the ring. The blue box surrounds the position of the knot. 
 \label{fig:NICI_annot}}
\end{figure}

The NICI images clearly reveal the non axi-symmetric shape of the
cavity. This is the highlight of our observations. The shape is
roughly elliptical, albeit with significant departures, with a
representative eccentricity of $e_c = \sqrt{1-(b_c/a_c)^2} \approx
0.5$ (where $a_c$ and $b_c$ are the ellipse axes). But, as illustrated
in Fig.~\ref{fig:NICI_annot}, the exterior ring can be described as
being composed of four arms overlapping in their extremities, except
for two small, $\sim$0.2~arcsec gaps or intensity nulls along the
ring, due north and south-south-east from the star. The observation of
these morphological feature is based on the shape of the inner rim of
the ring seen in direct images (even without subtraction of the PSF
glare).  Additionally, an unresolved brightness increment, or knot,
can be seen along the north-western segment (which is also the
brightest segment) - the alignment of this unresolved feature in all
filters is illustrated in the RGB image, although we cannot determine
from these data alone if this knot is real or perhaps only a ghost or
a chance alignment of speckles.

%{\bf FMe: Comment: Speckle would not be at the same place in Kband and L band. 
%In the RGB image, inside the gap you have R, G, and B dots, but no pink ones! 
%The speckle pattern should be homothetic with lambda. Static aberations on the contrary 
%may appear at the same place... (ghosts for example)}
%

The south-western spiral arm segment (labelled arm 2 on
Fig.~\ref{fig:NICI_annot}) sprouts away from the outer ring at a PA of
-45~deg (East of North), where it extends into the outer arm seen by
\citet{2006ApJ...636L.153F}. This outer arm is at much fainter levels
than the outer ring - our dataset has finer resolution but is
shallower, so that only its root is seen in our figures. As discussed
in \citet{2006ApJ...636L.153F}, this outer arm could have been
triggered by a stellar encounter, but the putative partner remains to
be identified. A nearby point source turned out to be a background
star \citep{2010PASJ...62..347F} - in our NICI images this source is
located at a separation of 5.\arcsec44 from HD~142527, and a PA of
220~deg (East of North).

%sep 5.43733387548145 PA -140.749477770614 
%pc-22-48-47-19013:30:33~/common/NICI/selfpsfsub$ perl HD142527B.pl
%sep 5.43733387548145 PA 39.2505222293865 

The inner radius of the ring varies from 0.7$^{\prime\prime}$ to the
north, to about 1.0$^{\prime\prime}$ to the south, and a projection
effect is discarded since the central star is clearly offset by
$0.17\pm0.02^{\prime\prime}$ from the centre of this approximate
ellipse. This offset was first noted by \citet{2006ApJ...636L.153F},
albeit at lower angular resolution.

In $K$-band the west side of the rim we see is brighter, in agreement
with \citet{2006ApJ...636L.153F}.  \citet{2006ApJ...644L.133F} and
\citet{2011A&A...528A..91V} already noted that in the thermal~IR, $N$
and $Q$ bands, it is the {\em eastern} side that is brightest, which
is consistent with a projection effect
\citep[][]{2006ApJ...644L.133F}. At the shorter wavelength $K$-band,
the front (i.e. nearest) side of the disk appears brighter because of forward
scattering at the surface of the disk. On the opposite, the thermal
emission from the larger projected surface of the rim of the back side
of the disk appears brighter than the slimmer (in projection) rim of
the near side. This implies a major axis of the disk in the
North-South direction and a minor axis in the E-W direction (as
observed here).  The emission null we detect due North is seen on both
sets of Subaru images, at $K$-band and in the mid-IR as well.

\subsection{Detection limits in the range 14--35 AU}

\begin{figure}
\begin{center}
\includegraphics[width=0.9\textwidth,height=!]{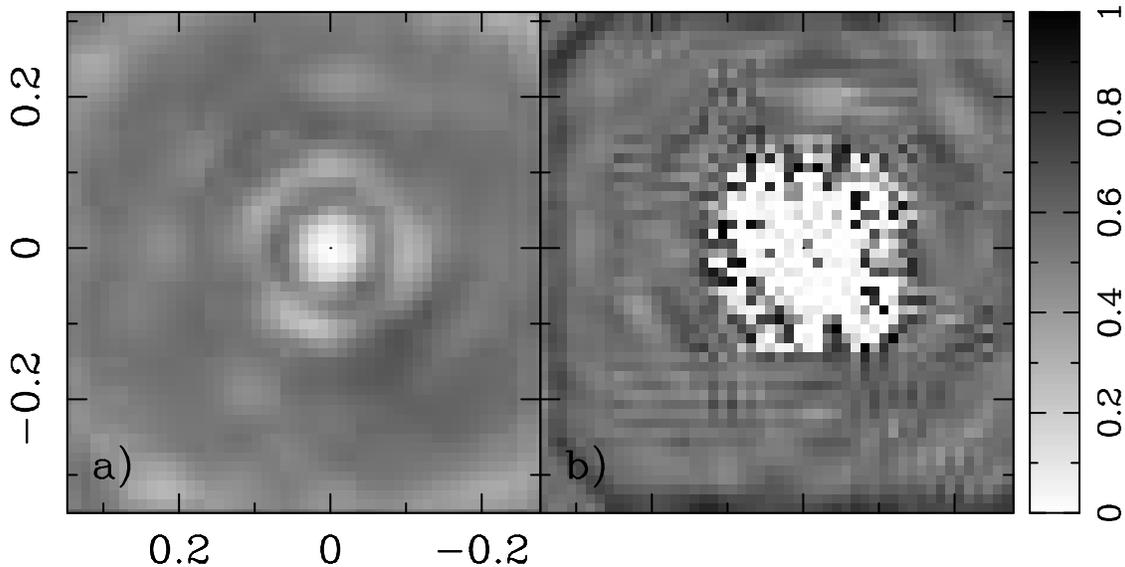}
\end{center}
\caption{Upper limits on the $K_s$ magnitude of HD~142527b. Axis
  labels are as for Fig.~\ref{fig:NICI}.  In a) (left panel) we show
  the collapsed SINFONI datacube, as produced by the pipeline, but
  divided by an azimuthal average chosen to highlight the PSF
  pattern. In b) (right panel) we show the collapsed image of the
  spectrally-deconvolved and PSF-subtracted data cube, with a linear
  intensity scale streching from -0.75 to 0.57, in counts, where the
  maximum number of counts in the collapsed cube (shown in a)) is
  548. The root-mean-square dispersion of the counts in an annulus
  from 0.2 to 0.3~arcsec is $9.6~10^{-2}$, or a 3~$\sigma$ lower limit
  to the flux ratio of 1905. Using a separate PSF standard, this
  residual rms value is 0.19, giving a 3~$\sigma$ lower limit to the
  flux ratio of 966.
 \label{fig:SINFO}}
\end{figure}

No companion is clearly detected in our NICI images. An upper limit on
a binary companion is difficult to obtain because of the speckle
noise, and because of the lack of accurate NICI zero points. However,
the SINFONI observations allow us to place a lower limit flux ratio at
0.2--0.3$^{\prime\prime}$ separation of 1905 at 3~$\sigma$ in $K$;
Fig.~\ref{fig:SINFO} illustrates this upper limit. Modeling the SED
\citep[with MCFOST,][]{2006A&A...459..797P} we estimate that half of
the total flux density of HD~142527 in K can be assigned to the
disk. In that case, the SINFONI flux ratio limit of 1905 corresponds
to an absolute magnitude $M_{K} > 8.2$ at 3~$\sigma$.  As a comparison
point we refer to the {\tt COND03} tracks \citep{2003A&A...402..701B}
at 1~Myr of age, which would imply a mass of less than
12~M$_\mathrm{jup}$ given this $M_{K}$ limit.  However the age of
HD~142527 and the models are not precise enough to ascertain this
upper limit mass value. We stress again that this limit constrains the
presence of massive gaseous giants at stellocentric radii of 14 to
35~AU.
% AJ: we use the COND03 tracks at 1 Myr but quote an age of 2 Myr before.
%     by using the lower age we are biasing our upper mass limit low (right?), in which case
%  we should use an age in the COND03 tracks >= 2 Myr to be conservative, or interpolate

% see ~/common/HD142527/selfpsf.pl  diskfig_sinfo_nocomp.pl fluxratios.pl fluxratios_nocomp.pl 

\section{Analysis} \label{sec:analysis}

%The total mass that has been cleared out of the cavity is about 
%$\sim$401~M$_\mathrm{jup}$. This can be estimated by extrapolating the
%peak surface density of $1.1~10^{-5}$M$_\odot$~AU$^{-2}$ at the inner
%edge of the outer disk, and subtracting the mass accreted by the star
%in 2~Myr\footnote{this estimate assumes that the the surface density
%  extends roughly with a inverse-radius law from $\sim 130~$AU to
%  $\sim~200$~AU, and that the total outer disk mass is
%  $\sim$0.1~M$_\odot$}.  
%

%\section{Discussion}

\subsection{Arguments in favor of a planet-created gap for HD~142527}

%~/common/HD142527_paper/gapmass.pl

The contribution from photoevaporation to cavity clearing can be
neglected in this case. The inner disk, the large gap and the vigorous
accretion rate \citep{2006A&A...459..837G} of $7 \times
10^{-8}$~M$_{\odot}$~yr~$^{-1}$ argue against photoevaporation.  An
X-ray driven wind can indeed explain a subset of the observed
accreting disks with inner holes, but only those with radii smaller
than 20~AU and accretion rates less than
10$^{-8}$~M$_{\odot}$~yr~$^{-1}$ \citep{2011MNRAS.412...13O}.

Gap clearing by single or multiple planets could potentially explain
wide gaps observed in transitional disks
\citep{2003MNRAS.342...79R}. We thus focus on the simplest possible
scenario to interpret the morphology of the cavity in HD~142527's
disk, i.e. dynamical clearing by one single embedded and
massive-Jupiter-size planet opening a gap in its gaseous disk.  The
maximum gap opening width will depend on the mass and the orbital
parameters of this planet \citep{2011ApJ...738..131D}. The cavity
forms due to the torque exerted by the star--planet binary
\citep{Artymowicz:1994eu}.  Elliptical cavities have been predicted
with $e_{\rm c} \approx 0.25$ in the case of {\it circular} binaries
with mass ratios $> 3\times10^{-3}$, in which case the gap becomes
large enough to deplete the region in which the eccentricity-damping
outer 1:3 Lindblad resonance is produced \citep{2006A&A...447..369K}.
Further numerical work \citep{Hosseinbor:2007eg} shows that, for
eccentric binaries, the outer edge of the cavity becomes even more
eccentric.  The observed eccentric cavity is then a recurrent feature
of gap-clearing models.

%Previous modelling of giant planets embedded in gaseous disks shows
%\citep{2006A&A...447..369K,Hosseinbor:2007eg} that the outer radius of
%the cavity $R_{\rm out}$ is located at roughly twice the semi-major
%axis of the binary $a$.  The measured average value $R_{\rm out}
%\approx 140\,$AU then implies $a \sim 70\,$AU.
%
%An intermediate binary eccentricity is not unexpected, and has in fact
%been predicted as a result of the disc--binary interaction both for
%planetary systems \citep{Goldreich:1980dt} and comparable-mass
%binaries \citep{Rodig:2011uq}.

\subsection{Fargo hydrodynamic simulations of gap clearing by a planet}

In order to test further the single-planet origin for the wide gap in
the gaseous disk of HD~142527, we conducted hydrodynamic simulations
with FARGO.  Our goal here is not to fit the data but to verify
qualitatively the tenability of this scenario for HD~142527.  FARGO is
a dedicated grid-based 2D code publicly available on the web
\citep{2000A&AS..141..165M}, and specifically designed for
planet--disk interactions.  The run presented in this paper
corresponds to a system evolution equivalent to 1.7~Myr.

%The simulations were carried out using
%FARGO's parallel version running on a 12 core unit.  

%In order to attain this age in a reasonable amount of computing time
%we specified a coarse time interval sampling (20 time steps per
%planet's orbit), a fairly standard spatial resolution (see below) and
%output writing to a minimum.
%

We picked common fiducial parameters for the model, as detailed
below. The model disk is initially axisymmetric, its surface density
is initially a power law of radius, $\Sigma = \Sigma_0 (r/r_0)^{-1}$,
where $\Sigma_0 = 566 \mathrm{g~cm}^{-2}$ ($6.3\times 10^{-5}$
M$_\odot$~AU$^{-2}$) at $r_0 = 1$~AU, so slightly smaller than the
value assumed by \citet{2011ApJ...738..131D}. The model disk has a
constant pressure scale height $H$ over radius $r$ with value
$H/r=0.05$.  For the disk calculations the boundaries are 4 and 300~AU
for the inner and outer radii, respectively, giving a total mass of
0.1~M$_\odot$. The inner boundary is open, allowing mass to fall onto
the star, while the outer boundary is such that we allow for a steady
state mass transfer into the disk. Accretion across the disk is
modeled by using an $\alpha$ prescription
\citep{1973A&A....24..337S}. We adopted $\alpha = 0.002$ following
\citet{2011ApJ...738..131D}. From the SINFONI limits described above
we know that any planet present in the inner 14 to 35~AU cannot have a
mass exceeding $\sim$12 Jupiter masses.  We assume no such planet is
present further out in the disk either, and chose a planet mass of
10~$M_{\rm Jup}$.

%\footnote{using the outer mass input
%  parameter in FARGO; the disk mass is kept constant}

%We have assumed that accretion onto the planet is negligible in the
%gap clearing process. Planet accretion is approximated by assuming
%that a fraction of the mass entering the planetary Hill radius is
%arbitrarily assigned to planet accretion. For a massive planet located
%tens of AUs away from the central star, the Hill radius is too large
%compared to the planet's radius for FARGO's approximation to be
%valid. Adaptive meshes are required to treat planet accretion
%realistically. In this hydrodynamic model we do not attempt to fit the
%accretion rates onto the star either.

We placed a 10~$M_{\rm Jup}$ planet in a fixed circular orbit at 90~AU
in the model disk. This radius was chosen so that the planet's sphere
of influence, given by a few times its Hill radius
\citep[e.g.][]{2011ApJ...738..131D}, reaches the observed gap size of
$\sim$130~AU. Migration was inhibited until a sizeable gap was carved
by the planet. This artificial constraint stems from the lack of heat
or momentum diffusion in FARGO \citep{2011ApJ...738..131D}.

\begin{figure}
\begin{center}
\includegraphics[width=0.9\textwidth,height=!]{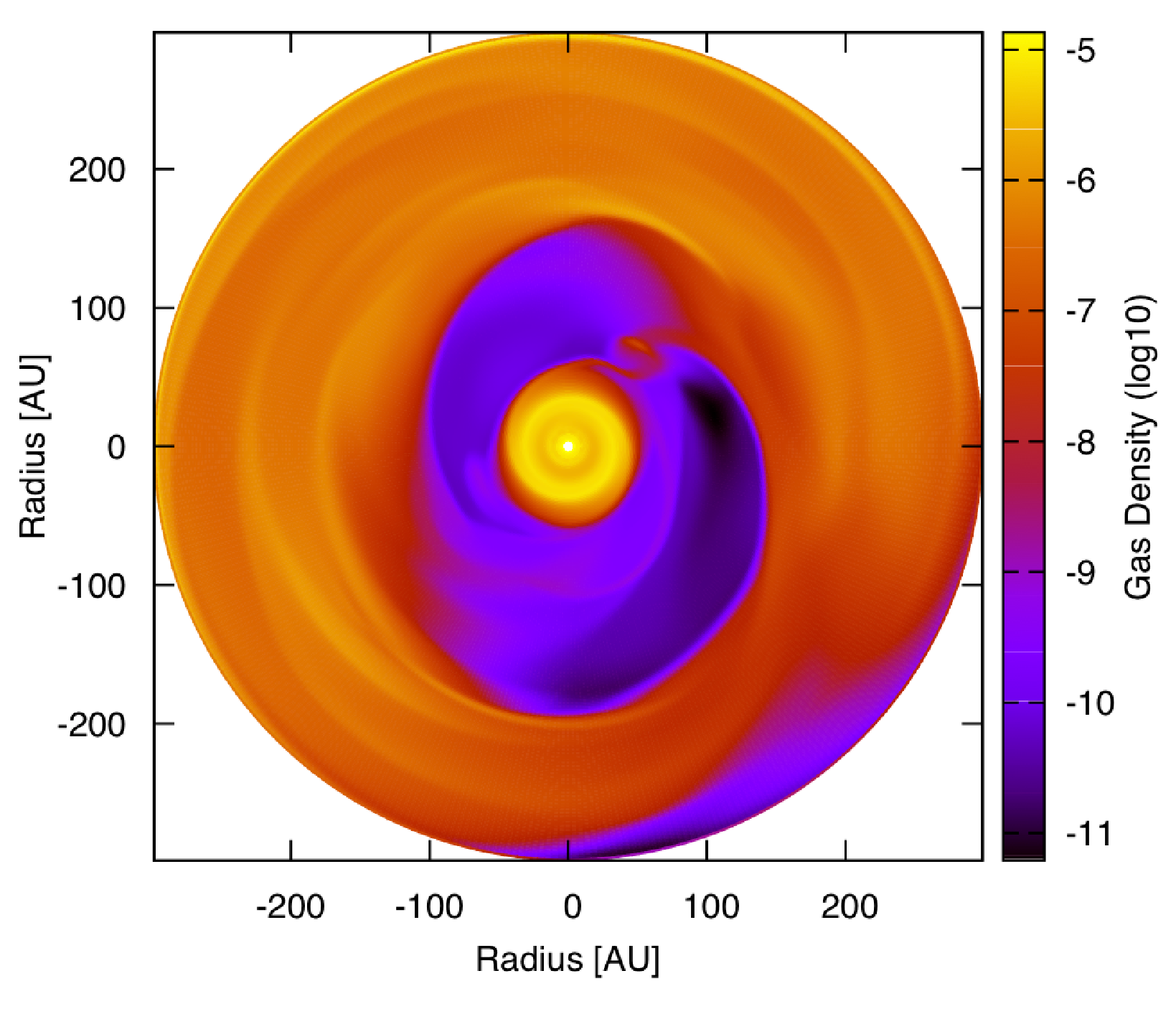}
\end{center}
\caption{FARGO simulation of a 10~M$_\mathrm{jup}$ planet embedded in
  a 0.1~M$_\odot$ disk. The spatial axes are in AU. The color scale
  represents the surface density in a logarithmic ramp in FARGO code units
  in which 10$^{-5}$ equals 88~g~cm$^{-2}$.  Notice the stellar shift
  relative to the position of the cavity centroid of $\sim$16~AU.
  \label{fig:fargo}}
\end{figure}

The resulting gas surface density field is shown in
Fig.~\ref{fig:fargo}. This specific dynamically perturbed morphology
will persist over a timescale of a few tens of thousand years (i.e,
around 100 orbits) but is not in steady state. The rim will remain
asymmetric over time, and is composed of tightly wound spiral arms,
whose superposition forms a roughly elliptical cavity. The cavity
centroid is persistently offset from the star, but any particular
morphology of the gap is transient and lasts about 100 orbits. In this
snapshot view (fig.~\ref{fig:fargo}), the outer edges of the cavity
are most reminiscent of those seen in the NICI images. The cavity
eccentricity is $\sim$0.6, and its centroid is offset from the star by
$\sim$16 AU.

%On the hill radius discussion. The way fargo deals with accretion
%onto planets is just by assuming that a fraction of the mass that
%enters the hill radius is accreted. This fraction is a parameter
%that you set by hand in the parameter file (a number between 0
%and 1). If i set it to 0.5, then half the mass that enters the
%Hill radius will be accreted, and so on.  Just assuming this and
%ignoring the creation of circumplanetary disks feels very
%artificial to me. So I also think that ignoring planet accretion
%is the way to proceed in this case.
%

As seen in Fig.~\ref{fig:fargo}, this model produces a massive inner
disc, extending out to 50-60~AU in radius, which is not consistent
with the observations.  Additional ingredients are required to deplete
that region in the framework of a dynamical clearing model, e.g.,
additional planets closer to the star \citep[as
  in][]{2011ApJ...738..131D}. Fitting the inner disc is left for
future work.  We stress that the purpose of these simulations is to
provide a plausible explanation of the NICI observations of the inner
edge of the outer disk only.

%We do
%not aim to reproduce the properties of the inner disk, which is also
%modelled in the FARGO simulations. In fact, the simulations produce an
%inner disk that is much too large and massive compared to the
%available data.
%
%This mismatch could be due to
%the limitations of our simple model, or a possible indication for the
%existence of multiple planets.
%

% AJ: references for this last statement?

\section{Summary and discussion} \label{sec:conclusion}

Our observations of HD~142527 indicate that the inner rim of its outer
disk is roughly elliptical, and composed of an overlapping set of
spiral arm segments. We confirm that the star is offset by
$0.17\pm0.02^{\prime\prime}$ from the centroid of the cavity. No
companion is detected in the range of 14 to 35~AU, with a lower limit
$M_{K} > 8.2$ at 3~$\sigma$ (for a distance of 140~pc).

The observed morphology of the disk in HD~142527 suggests a
dynamically perturbed state. The segmented spirals, and the offset
between the cavity centroid and the star, can be seen among the varied
morphologies predicted by hydrodynamic simulations of HD~142527.  To
explore, at least qualitatively, the origin of the cavity and its
peculiar shape, we reported on a simple case.  A massive protoplanet
on a circular orbit whose radius is comparable to the disk's outer
edge perturbs the entire disk, which does not reach steady state even
after thousands of orbits. The global shape of the cavity is a
long-standing feature; the general asymmetries (off-center star,
eccentricity of the cavity) are imprinted early and remain for the
duration of the calculations.  However, the specific configuration of
the model run, position of the streamers, and position and number of
spiral arms, form a specific morphology that is only a transient
phase. A particular morphology may recur, but lasts for a limited
number of orbits.  That phase which qualitatively matches HD~142527
lasts for a hundred orbits or so.

It is tempting to suggest that the observed shape of the disk rim of
HD~142527 today will also be brief compared to its age, with features
that are unlikely to be found in other systems that are also
undergoing dynamical clearing. In this scenario, snapshot imaging of
transition disks are likely to look different in their fine details.
However, there should be common features like off-centering of the
central star, non axi-symmetry of the disk rim, and presence of spiral
arms. Current imaging campaigns are indeed revealing these structures,
as in HD~100546 \citep[][]{Grady2001}; HD~135344B
\citep[][]{2012ApJ...748L..22M}; AB~Aur
\citep[][]{2011ApJ...729L..17H}. Interstingly, the list of known
transition disks is rapdily increasing with improving observational
capabilities both in direct imaging at optical/near-IR (see e.g., the
citations above) and in the sub-mm regime
\citep[][]{Andrews2011}. We should be able to verify these
ideas in the near future, specifically that the observed morphological
variety of transition disks derives from a common dynamical history in
this evolutionnary stage when there is large-scale feedback between
the planet formation process and its parent disk.

%We propose that this particular transient state of HD~142527 will be
%brief compared to its age, with features that are unlikely to be found
%in other systems that are undergoing dynamical clearing. In this
%scenario snapshots of such transition disks are bound to be
%unique. 

%An interesting comparison object is HD~135344B, where
%\citet{2012ApJ...748L..22M} also found spiral structure in the outer
%disk - but with a much smaller cavity and a very different spiral
%structure. Another comparison object is AB~Aur, where
%\citet{2011ApJ...729L..17H} find a large gap at $\sim$80~AU, with
%intensity decrements along the outer ring and several spiral arms
%sprouting outwards.
%

A key to test this transient scenario for HD~142527 and transition
disks in general would be to detect the gap-crossing planetary
accretion flow and link it to the outer-disk asymmetries. Meanwhile, a
quantitative comparison with hydrodynamic models requires radiative
transfer calculations on the model density fields.

While this article was under review \citet{Biller2012} reported on a
tentative detection of a low-mass stellar companion to HD~142527, at
$\sim$13~AU, so abutting on the inner disk. This companion is inferred
through parametrised-modeling of visibility data. If real, this
stellar companion will strongly impact on the inner disk, and its
inclusion in simulation may bring closer agreement with the observed
inner disk properties.  This would be particularly interesting given
that our one-planet scenario currently fails to reproduce the
properties of the inner disk. It is currently too large in radius and
mass, compared with observations. Including a second body in the
simulation would produce a larger gap with a smaller inner disk
\citep[e.g,][]{2011ApJ...738..131D}.  However, HD~142527B alone
probably cannot explain the extent of the cavity. A single body cannot
clear a gap wider than $\sim$5 times its tidal radius
\citep[e.g.][]{2011ApJ...738..131D}, given by $R_{\rm hill}$, where
$R_{\rm hill} = a ( M_{p}/3M_{\star})^{1/3}$ is known as the Hill
radius.  Mass and semi-major axis estimates for the companion star
detected by \citet{Biller2012} give a Hill radius $R_{\rm hill} \sim
3-5$~AU.

\acknowledgments

SC, SP, AJ, FM, MS and AH acknowledge support from the Millennium
Science Initiative (Chilean Ministry of Economy), through grant
``Nucleus P10-022-F''. We thank the ESO Director General for
discretionary time that allowed the K-band limit. SC, AJ and JC thank
the CATA (``Fondo Basal PFB-06, CONICYT''). SC acknowledges FONDECYT
grant 1100221. AJ acknowledges additional support from Anillo ACT-086
and FONDECYT project 1095213.  FM acknowledges support from PNPS of
CNRS/INSU and from EU FP7-2011 under grant agreement No. 284405
(DIANA). JC acknowledges VRI-PUC (Inicio 16/2010) and FONDECYT
11100240.

\label{lastpage}

\end{document}